
\documentstyle{amsppt} 
\pagewidth{18cm}\pageheight{23cm}
\CenteredTagsOnSplits
\rightheadtext{Problem of Metrizability\dots}
\topmatter
\title
Problem of Metrizability for the Dynamical Systems\\
Accepting the Normal Shift.
\endtitle
\author
Sharipov R.A.
\endauthor
\address
Department of Mathematics,
Bashkir State University, Frunze str. 32, 450074 Ufa,
Russia
\endaddress
\email
root\@bgua.bashkiria.su
\endemail
\date
June, 1993.
\enddate
\abstract
     The problem of metrizability for the dynamical systems
accepting the normal shift is formulated and solved. The explicit
formula for the force field of metrizable Newtonian dynamical
system $\ddot\bold r=\bold F(\bold r,\dot\bold r)$ is found.
\endabstract
\endtopmatter
\document
\head
1. Introduction.
\endhead
     The class of Newtonian dynamical systems accepting the
normal shift was first defined in \cite{1} (see also \cite{2} and
\cite{3}). It's the class of dynamical systems in $\Bbb R^n$
given by the differential equations of the form
$$
\ddot\bold r=\bold F(\bold r,\dot\bold r)\tag1.1
$$
and possessing some additional geometrical property: the property
of conserving the orthogonality of trajectories and the
hypersurfaces shifted along these trajectories. The idea for
considering such systems was found as a result of generalizing
the classical construction of normal shift which is known also
as a Bonnet transformation. Let S be the hypersurface in
$\Bbb R^n$. From each point $M$ on $S$ we draw the segment $MM'$
with the fixed length $l$ along the normal vector to $S$. The
points $M'$ then form another hypersurface $S'$, the segment
$MM'$ being perpendicular to $S'$. The transformation
$f:S\longrightarrow S'$ just described is the classical Bonnet
transformation. It has the generalization for non-Euclidean
situation: for the Riemannian metric $g_{ij}$ one should replace
the segment $MM'$ of a straight line by the segment of geodesic
line with the length $l$. The transformation $f:S\longrightarrow
S'$ in this case is the metrical Bonnet transformation or the
normal shift with respect to the metric $g_{ij}$. Trajectories of
this shift are defined by the equation of geodesic lines
$$
\ddot r^k=-\Gamma^k_{ij}\dot r^i\dot r^j\tag1.2
$$
This equation can be treated as a particular case of the
Newtonian dynamical system (same indices on different levels in
\thetag{1.2} and everywhere below imply summation). \par
     In \cite{1} (see also \cite{2} and \cite{3}) the classical
Bonnet transformation was generalized for the case of dynamical
systems \thetag{1.1} in $\Bbb R^n$. Other generalization is
a metrical Bonnet transformation. So quite natural question is:
how do these two generalizations relate each other? In order to
treat \thetag{1.2} as a dynamical system accepting the normal
shift in standard Euclidean metric in $\Bbb R^n$ we should have
the coincidence of the concept of orthogonality with respect to
both Euclidean and non-Euclidean metrics $\delta_{ij}$ and
$g_{ij}$ in $\Bbb R^n$. That is $g_{ij}=e^{-2f}\delta_{ij}$
should be conformally Euclidean metric. Problem of metrizability
then can be stated as follows.
\proclaim{\quad Problem of metrizability} Under which
circumstances the dynamical system \thetag{1.1} accepting the
normal shift in the sense of \cite{1} and \cite{4} is equivalent
to the metrical normal shift for some conformally Euclidean
metric in $\Bbb R^n$.
\endproclaim
Studying this problem and solving it is the goal of present
paper. As we shall see below its solution is explicit and
constructive. \par
\head
2. Geodesic lines of conformally Euclidean metric in $\Bbb R^n$
\endhead
     Let $\delta_{ij}$ be the metric tensor for the standard
Euclidean metric in $\Bbb R^n$. Let's consider conformally
Euclidean metric
$$
g_{ij}=e^{-2f}\delta_{ij}\tag2.1
$$
where $f=f(\bold r)=f(r^1,\dots,r^n)$ is some scalar function in
$\Bbb R^n$. Metrical connection for the metric \thetag{2.1} is
given by
$$
\Gamma^k_{ij}=\delta_{ij}\delta^{ks}\partial_sf-
(\delta^k_i\partial_jf+\delta^k_j\partial_if)\tag2.2
$$
Here by $\partial_s$ we denote the partial derivative with
respect to $r^s$. The equation of geodesic lines for \thetag{2.2}
has the following form
$$
\ddot r^k=-\delta^{ks}\partial_sf\dot r^i \delta_{ij}\dot r^j+
2\partial_if\dot r^i\dot r^k\tag2.3
$$
In a vectorial form like \thetag{1.1} the force field for the
dynamical system corresponding to \thetag{2.3} is written as
follows
$$
\bold F(\bold r,\bold v)=-\nabla f |\bold v|^2+2\left<\nabla f,
\bold v\right> \bold v\tag2.4
$$
Here $\nabla f$ is a gradient of the function $f$ considered as a
vector in $\Bbb R^n$ and $\left<\nabla f,\bold v\right>$ is a
standard Euclidean scalar product of $\nabla f$ with the vector
of velocity $\bold v$. The modulus of the velocity vector in
\thetag{2.4} is also calculated in a standard metric
$\delta_{ij}$. \par
     Because of its geometrical origin the dynamical system
\thetag{1.1} with force field \thetag{2.4} accepts the normal
shift. However it is curious to check this fact directly.
Following the receipt of \cite{4} and \cite{5} we choose the
coordinates $u^1,\dots,u^{n-1}$ on a unit sphere $|\bold v|=1$ in
the velocity space. Let's denote $v=|\bold v|$ and let $\bold N$
be the unit vector directed along the vector $\bold v$. Then
$$
\bold N=\bold N(u^1,\dots,u^{n-1})\hskip 5em
\bold M_i=\frac{\partial\bold N}{\partial u^i}\tag2.5
$$
For the derivatives of the vectors $\bold M_i$ defined by
\thetag{2.5} one has a Weingarten derivation formula
$$
\frac{\partial\bold M_i}{\partial u^j}=\vartheta^k_{ij}
\bold M_k-G_{ij}\bold N\tag2.6
$$
where $G_{ij}=\left<\bold M_i,\bold M_j\right>$ is an induced
metric on the unit sphere. Let's consider the expansion of the
force field \thetag{2.4} in the basis formed by $\bold N$ and
the vectors $\bold M_i$
$$
\bold F=A\bold N+B^i\bold M_i\tag2.7
$$
Let's find the spatial gradients of the coefficients of the
expansion \thetag{2.7} and then let's expand them in the same
basis. As a result we get
$$
\frac{\partial A}{\partial r^k}=a N_k+\alpha^pM_{pk}\qquad
\frac{\partial B^i}{\partial r^k}=b^i N_k+\beta^{ip}M_{pk}
\tag2.8
$$
According to the results of \cite{4} and \cite{5} the weak
normality condition for the dynamical system with the force field
\thetag{2.7} is given by the following equations
$$
\align
B^i&=-G^{ik}\frac{\partial A}{\partial u^k}\tag2.9 \\
\split
\alpha^i&+\frac{B^qB^k}{v^2}\vartheta^i_{qk}-\frac{B^i A}{v^2}+
b^i+\\ &+\frac{A}{v}\frac{\partial B^i}{\partial v}+
\frac{\partial B^i}{\partial u^k}\frac{B^k}{v^2}-
\frac{b^i}{v}\frac{\partial A}{\partial v}=0
\endsplit\tag2.10
\endalign
$$
In order to check the equations \thetag{2.9} and \thetag{2.10}
in the case of force field \thetag{2.4} we calculate the
coefficients of expansion \thetag{2.7} explicitly. Because of
the orthogonality $\left<\bold N,\bold M_i\right>=0$ we have
$$
\aligned
A&=\left<\bold F,\bold N\right>=\left<\nabla f,\bold N
\right> v^2\\ B^i&=G^{ik}\left<\bold F,\bold M_k\right>=
-G^{ik}\left<\nabla f,\bold M_k\right> v^2
\endaligned
\tag2.11
$$
{}From \thetag{2.11} we can see that the relationship \thetag{2.9}
becomes the identity due to \thetag{2.5}. For the coefficients
$\alpha^i$ and $b^i$ in \thetag{2.8} we find
$$
\align
\alpha^i&=G^{ik}\frac{\partial A}{\partial r^q}\delta^{qs}
M_{ks}=G^{ik}\frac{\partial^2f}{\partial r^q\partial r^m}
\delta^{mp}N_p\delta^{qs}M_{ks}v^2 \\
b^i&=\frac{\partial B^i}{\partial r^q}\delta^{qs} N_s =
-G^{ik}\frac{\partial^2f}{\partial r^q\partial r^m}
\delta^{mp}M_{kp}\delta^{qs}N_sv^2
\endalign
$$
{}From these two equalities we see that $\alpha^i$ and $b^i$ differ
only by sign. When substituting them into \thetag{2.10} they
vanish. Let's calculate the sixth term in \thetag{2.10} using
the formula \thetag{2.6}
$$
\frac{\partial B^i}{\partial u^k}\frac{B^k}{v^2}=
\frac{B^iA}{v^2}+\left(\frac{\partial G^{iq}}{\partial u^k}+
G^{is}\vartheta^q_{sk}\right) G_{qp}\frac{B^pB^k}{v^2}
$$
Taking into account the concordance of the metric $G_{ij}$ and
the metrical connection $\vartheta^k_{ij}$ we can bring this
equation into the following form
$$
\frac{\partial B^i}{\partial u^k}\frac{B^k}{v^2}=
\frac{B^iA}{v^2}-G^{sq}\vartheta^i_{sk}G_{qp}
\frac{B^pB^k}{v^2}=\frac{B^iA}{v^2}-\frac{B^sB^k}{v^2}
\vartheta^i_{sk}
$$
When substituting this form of sixth term into \thetag{2.10} it
cancels the second and the third terms in \thetag{2.10}.
Because of quadratic dependence of $A$ and $B^i$ in \thetag{2.11}
upon $v$ fifth and seventh terms in \thetag{2.10} cancel each
other. Resuming all above we conclude that the equations
\thetag{2.9} and \thetag{2.10} hold identically for the
components of the expansion \thetag{2.7}. Weak normality
condition for \thetag{2.4} is fulfilled. \par
     Next step consists in substituting the geodesic flows of the
form \thetag{2.4} by some dynamical systems similar to them.
Let's start with the Euclidean metric $g_{ij}=\delta_{ij}$. Force
field \thetag{2.4} then is zero $\bold F(\bold r,\bold v)=0$,
trajectories are straight lines. For the dynamical system with
the force field $\bold F(\bold r,\bold v)=\bold v$ they are also
straight lines. From geometrical point of view these two
dynamical systems realize the same normal shift. This example
shows that for to solve the problem of metrizability one should
find all dynamical systems accepting the normal shift in
$\Bbb R^n$ for which the trajectories are the geodesic lines of
conformally Euclidean metrics. The problem of geometrical
coincidence of trajectories for two different dynamical systems
was first stated in \cite{6}. There the following pairs of
dynamical systems were considered
$$
\aligned
&\ddot r^k+\Gamma^k_{ij}\dot r^i\dot r^j=
F^k(\bold r,\dot\bold r)\\
&\ddot r^k+\tilde\Gamma^k_{ij}\dot r^i\dot r^j=
\tilde F^k(\bold r,\dot\bold r)
\endaligned
\tag2.12
$$
In the case of $F^k=\tilde F^k=0$ the condition of coincidence
of trajectories for the dynamical systems \thetag{2.12} is known
as a condition of geodesical equivalence for the affine
connections $\Gamma^k_{ij}$ and $\tilde\Gamma^k_{ij}$. The
detailed discussion of the questions connected with geodesical
equivalence and geodesical maps can be found in the monograph
\cite{7} (see also \cite{8} and \cite{9}). \par
     For nonzero $F^k$ and $\tilde F^k$ if the trajectories of
the dynamical systems \thetag{2.12} coincide then one say that
one of these systems is a modeling system for another. This
case was considered in \cite{10-13}. The terms {\it inheriting
the trajectories} and {\it trajectory equivalence} below seem to
be more preferable than the term {\it modeling} from purely
linguistical point of view. \par
\head
3. Inheriting the trajectories and trajectory equivalence
of dynamical systems.
\endhead
     Let's consider the pair of Newtonian dynamical systems
\thetag{1.1} of the second order
$$
\align
\partial_{tt}\bold r&=\bold F_1(\bold r,\partial_t\bold r)
\tag3.1\\
\partial_{\tau\tau}\bold r&=\bold F_2(\bold r,
\partial_\tau\bold r)\tag3.2
\endalign
$$
Trajectories of these systems are defined by the initial position
and initial velocity
$$
\align
&\left.\bold r\right|_{t=0}=\bold r_0\hskip 10em
\left.\partial_t\bold r\right|_{t=0}=\bold v_0\tag3.3\\
&\left.\bold r\right|_{\tau=0}=\bold r_0\hskip 10em
\left.\partial_\tau\bold r\right|_{\tau=0}=\bold w_0\tag3.4
\endalign
$$
Let $\bold r=\bold R_1(t,\bold r_0,\bold v_0)$ and
$\bold r=\bold R_2(\tau,\bold r_0,\bold w_0)$ are the solutions
of the equations \thetag{3.1} and \thetag{3.2} with the initial
conditions \thetag{3.3} and \thetag{3.4}. \par
\definition{Definition 1} Say that the dynamical system
\thetag{3.2} inherits the trajectories of the system \thetag{3.1}
if for any pair of vectors $\bold r_0$ and $\bold w_0\neq 0$ one
can find the vector $\bold v_0$ and the function $T(\tau)$ such
that $T(0)=0$ and the following equality
$$
\bold R_1(T(\tau),\bold r_0,\bold v_0)=
\bold R_2(\tau,\bold r_0,\bold w_0)\tag3.5
$$
holds identically by $\tau$ in some neighborhood of zero $\tau=
0$.
\enddefinition
\definition{Definition 2} Two dynamical systems are called
trajectory equivalent if each of them inherits the trajectories
of the other.
\enddefinition
     Differentiating \thetag{3.5} by $\tau$ for $\tau=0$ we get
the relation between the vectors $\bold w_0$ and $\bold v_0$ in
the following form
$$
\bold v_0\,\partial_\tau T(0)=\bold w_0\tag3.6
$$
{}From (3.6) we see that $\bold v_0\neq 0$ and $\partial_\tau
T(0)\neq 0$. For nonzero $\tau$ we have
$$
\partial_t\bold R_1(T(\tau),\bold r_0,\bold v_0)\,
\partial_\tau T=\partial_\tau \bold R_2(\tau,\bold r_0,
\bold w_0)\tag3.7
$$
Let's differentiate the relationship \thetag{3.7} by $\tau$.
Then for $\tau=0$ we get
$$
\bold F_1(\bold r_0,\bold v_0)\,\partial_\tau T(0)^2+
\bold v_0\,\partial_{\tau\tau} T(0)=
\bold F_2(\bold r_0,\bold w_0)\tag3.8
$$
The relationship \thetag{3.8} bind the force field of the
dynamical system \thetag{3.2} with the force field of the system
\thetag{3.1} trajectories of which are inherited according to
the definition 1. Because of $\bold v_0\neq\bold w_0$ this
relationship is nonlocal. However, if $\bold F_1(\bold r,
\bold v)$ is homogeneous function by $\bold v$ this relationship
becomes local. It's the very case we shall consider below. \par
     Let $\gamma$ be the degree of homogeneity for the function
$\bold F_1(\bold r,\bold v)$ with respect to its vectorial
argument $\bold v$. The relationship \thetag{3.8} then has the
following form
$$
\bold F_1(\bold r,\bold w)\,\partial_\tau T(0)^{2-\gamma}+
\bold w\,\partial_{\tau\tau}T(0)\,\partial_\tau T(0)^{-1}=
\bold F_2(\bold r,\bold w)\tag3.9
$$
Since the initial data $\bold r_0$ and $\bold w_0$ in
\thetag{3.4} are arbitrary we omit the index $0$ everywhere in
\thetag{3.9}. Because of \thetag{3.9} the vectors $\bold w$,
$\bold F_1(\bold r,\bold w)$ and $\bold F_2(\bold r,\bold w)$
are linearly dependent. Two cases are possible: \par
\roster
\item special case when the vectors $\bold F_1(\bold r,\bold w)$
and $\bold w$ are linearly dependent,
\item generic case when vectors $\bold F_1(\bold r,\bold w)$ and
$\bold w$ are linearly independent.
\endroster
     Let's start with the first case. Here for the force field of
the dynamical system \thetag{3.1} we have
$$
\bold F_1(\bold r,\bold w)=\bold w\frac{H_1(\bold r,\bold w)}
{|\bold w|}\tag3.10
$$
where $H_1(\bold r,\bold w)$ is a scalar function homogeneous
by $\bold w$ with order of homogeneity $\gamma$. Because of
\thetag{3.9} the force field $\bold F_2(\bold r,\bold w)$ has the
similar form
$$
\bold F_2(\bold r,\bold w)=\bold w\frac{H_2(\bold r,\bold w)}
{|\bold w|}\tag3.11
$$
However the function $H_2(\bold r,\bold w)$ in \thetag{3.11}
shouldn't be homogeneous by $\bold w$. Trajectories of the
dynamical systems \thetag{3.1} and \thetag{3.2} with the force
fields \thetag{3.10} and \thetag{3.11} are straight lines,
therefore any two such systems are inheriting the trajectories
of each other even when the function $H_1(\bold r,\bold w)$ is
not homogeneous by $\bold w$. \par
     The second case is more complicated. Here vector $\bold F_2
(\bold r,\bold w)$ can be decomposed by the vectors $\bold F_1
(\bold r,\bold w)$ and $\bold w$
$$
\bold F_2(\bold r,\bold w)=C(\bold r,\bold w)
\bold F_1(\bold r,\bold w)+\bold w
\frac{H(\bold r,\bold w)}{|\bold w|}\tag3.12
$$
Differentiating \thetag{3.7} by $\tau$ for $\tau\neq 0$ we get the
equation for the function $T(\tau)$
$$
\bold F_1(\bold r,\bold w)\,\partial_\tau T^{2-\gamma}+
\bold w\,\partial_{\tau\tau}T \,\partial_{\tau}T^{-1}=
\bold F_2(\bold r,\bold w)\tag3.13
$$
where $\bold r$ and $\bold w$ are the functions of $\tau$ defined
by \thetag{3.2} and \thetag{3.4}
$$
\bold r=\bold r(\tau)=\bold R_2(\tau,\bold r_0,\bold w_0)
\hskip 10em \bold w=\bold w(\tau)=\partial_\tau\bold r
$$
Since the expansion of $\bold F_2(\bold r,\bold w)$ by two
linearly independent vectors $\bold F_1(\bold r,\bold w)$ and
$\bold w$ is unique from \thetag{3.12} and \thetag{3.13} we get
$$
\aligned
&(\partial_\tau T)^{2-\gamma}=C(\bold r,\bold w) \\
&\partial_{\tau\tau} T=\frac{H(\bold r,\bold w)}{|\bold w|}
 \partial_\tau T
\endaligned\tag3.14
$$
Because of \thetag{3.14} we should consider two different
subcases depending on the value of $\gamma$: $\gamma\neq 2$ and
$\gamma=2$.
     Let's consider the first subcase. Here the function
$T(\tau)$ is defined by the equation of the first order
$\partial_\tau T=C(\bold r,\bold w)^{\gamma-2}$. Therefore the
second equation \thetag{3.14} should be the consequence of the
first one. Differentiating the first equation \thetag{3.14}
with respect to $\tau$ we get
$$
\partial_\tau\ln(\partial_\tau T)=(\gamma-2)
\left(w^i\frac{\partial\ln(C)}{\partial r^i}+
F^i_2\frac{\partial\ln(C)}{\partial w^i}\right)
$$
Comparing this equation with the second equation \thetag{3.14}
and substituting $F^i_2$ in it by \thetag{3.13} we get the
relationship
$$
\frac{H}{|\bold w|}=\left(w^i\frac{\partial\ln(C)}{\partial r^i}
+F^i_1\frac{\partial C}{\partial w^i}\right)
\left(\frac{1}{\gamma-2}-w^i\frac{\partial ln(C)}{\partial w^i}
\right)^{-1}
$$
which express $H(\bold r,\bold w)$ through $C(\bold r,\bold w)$.
Force field $\bold F_2(\bold r,\bold w)$ of the dynamical system
\thetag{3.2} inheriting the trajectories of the system
\thetag{3.1} in this case is defined by one arbitrary scalar
function $C(\bold r,\bold w)$. \par
     Now let's consider the second subcase $\gamma=2$ Here the
first equation \thetag{3.14} is trivial and $C(\bold r,\bold w)
=1$. The function $T(\tau)$ is defined only by the second
equation \thetag{3.14}. The force field of the system
\thetag{3.2} defined by \thetag{3.12} contains one arbitrary
function $H(\bold r,\bold w)$. This case $\gamma=2$ is the most
interesting since the force fields of the form \thetag{2.4} are
covered by this case. Dynamical systems inheriting their
trajectories are defined by the following force fields
$$
\bold F(\bold r,\bold v)=-\nabla f |\bold v|^2+2\left<\nabla f,
\bold v\right> \bold v +\frac{\bold v}{|\bold v|}
H(\bold r,\bold v)\tag3.15
$$
where $f=f(\bold r)$. However not all the dynamical systems with
the force field \thetag{3.15} are accepting the normal shift.
\par
\head
4. Problem of metrizability.
\endhead
     Let the dynamical system \thetag{1.1} with the force field
\thetag{3.15} be accepting the normal shift. The force field
\thetag{3.15} differs from \thetag{2.4} by the term $\bold N
H(\bold r,\bold v)$. Therefore the components of the expansion
\thetag{2.7} for \thetag{3.15} are slightly different from that
of \thetag{2.4}
$$
\tilde A=A+H\hskip 10em\tilde B^i=B^i\tag4.1
$$
But both they are satisfying the equations \thetag{2.9} and
\thetag{2.10}. Substituting \thetag{4.1} into \thetag{2.9} we get
the vanishing of the derivatives
$$
\frac{\partial H}{\partial u^k}=0\hskip 10em
H=H(\bold r,v)=H(\bold r,|\bold v|)
$$
Therefore the function $H$ in \thetag{3.15} doesn't depend on the
direction of the vector of velocity. It depends only on the
modulus of velocity and on the position of the mass point in the
space. For the components of the expansion \thetag{2.8} which
are used in \thetag{2.10} we have
$$
\tilde\alpha^i=\alpha^i+h^i\hskip 10em \tilde b^i=b^i\tag4.2
$$
The parameters $h^i$ are defined by the function $H(\bold r,v)$
according to the formula
$$
h^i=G^{ik}\left<\nabla H,\bold M_k\right>=G^{ik}
\frac{\partial H}{\partial r^q}\delta^{qs}M_{ks}\tag4.3
$$
Let's substitute \thetag{4.1} and \thetag{4.2} into the equation
\thetag{2.10}. As a result we have
$$
h^i-\frac{B^i H}{v^2}+\frac{H}{v}\frac{\partial B^i}{\partial v}
-\frac{B^i}{v}\frac{\partial H}{\partial v}=0 \tag4.4
$$
The relationships \thetag{4.4} form the system of $n-1$ equations
for the function $H(\bold r, v)$. Taking into account
\thetag{2.11} and \thetag{4.3} we can transform them to the
following ones
$$
\frac{\partial H}{\partial r^k}+\left(v \frac{\partial f}
{\partial r^k}\right)\frac{\partial H}{\partial v}=
\frac{\partial f}{\partial r^k} H \tag4.5
$$
Let's consider the vector fields which are defined by the
differential equations \thetag{4.5}
$$
X^k=\frac{\partial}{\partial r^k}+
\left(v\frac{\partial f}{\partial r^k}\right)
\frac{\partial}{\partial v}\tag4.6
$$
It's easy to check that the vector fields \thetag{4.6} are
commuting, therefore the equations \thetag{4.5} are compatible.
It's wonderful that the common solution for the system of
equations \thetag{4.5} can be written explicitly
$$
H(\bold r,v)=H(ve^{-f})e^f\tag4.7
$$
Using \thetag{4.7} the final result can be formulated as a
following theorem giving the solution for the problem of
metrizability. \par
\proclaim{Theorem 1} The dynamical system \thetag{1.1} in
$\Bbb R^n$ is accepting the normal shift if its force field
has the following form
$$
\bold F(\bold r,\bold v)=-\nabla f |\bold v|^2+2\left<\nabla f,
\bold v\right> \bold v + \frac{\bold v}{|\bold v|}
H(|\bold v|e^{-f})e^f
$$
where $f=f(\bold r)=f(r^1,\dots,r^n)$ and $H=H(v)$ are two
arbitrary functions. When it is metrizable the dynamical system
\thetag{1.1} realizes the metrical normal shift for some
conformally Euclidean metric in $\Bbb R^n$.
\endproclaim
     The comparison of the force field from the theorem 1 with
the examples of \cite{1} shows that even in $\Bbb R^2$ one can
find the non-metrizable dynamical systems accepting the normal
shift. So the concept of the dynamical systems accepting the
normal shift is more wide, it cannot be reduced to the metrical
normal shift.

\enddocument